\documentclass[preprintnumbers, prd, floatfix,twocolumn,preprintnumbers, letterpaper, superscriptaddress,nofootinbib]{revtex4-1}
\usepackage{amsfonts}
\usepackage{amsmath}
\usepackage{amssymb,epsf}
\usepackage{latexsym}
\usepackage{graphicx,epsfig}
\usepackage{amssymb}
\usepackage{float}
\usepackage{subfigure}
\usepackage{epstopdf}
\usepackage[colorlinks=true,citecolor=blue,linkcolor=blue,urlcolor=black]{hyperref}
\usepackage{dcolumn}
\usepackage{psfrag}
\usepackage{wrapfig}
\usepackage{makeidx}
\usepackage{epsf}
\usepackage{color}
\usepackage{multirow}
\usepackage{mathtools}
\usepackage[normalem]{ulem}

\DeclarePairedDelimiter\abs{\lvert}{\rvert}
\graphicspath{{Images/}}

\usepackage{tikz}

\definecolor{lime}{HTML}{A6CE39}
\newcommand{\orcidicon}{%
    \begin{tikzpicture}
    \draw[lime, fill=lime] (0,0)
        circle [radius=0.16]
        node[white] {{\fontfamily{qag}\selectfont \tiny ID}};
    \draw[white, fill=white] (-0.0625,0.095)
        circle [radius=0.007];
    \end{tikzpicture}   \hspace{-2mm}
}
\newcommand\orcidFrancisco{{\href{https://orcid.org/0000-0002-9388-8373}{\orcidicon}}}
\newcommand\orcidMahdi{{\href{https://orcid.org/0000-0003-1196-9493}{\orcidicon}}}
\newcommand\orcidHooman{{\href{https://orcid.org/0000-0003-0941-8422}{\orcidicon}}}

\begin{document}

\title{Evolving traversable wormholes satisfying the energy conditions\\ in the presence of pole dark energy}
\author{Mahdi Kord Zangeneh\orcidMahdi\!\!}
\email{mkzangeneh@scu.ac.ir}
\affiliation{Physics Department, Faculty of Science, Shahid Chamran University of Ahvaz,
Ahvaz 61357-43135, Iran}
\author{Francisco S. N. Lobo\orcidFrancisco\!\!}
\email{fslobo@fc.ul.pt}
\affiliation{Instituto de Astrof\'isica e Ci\^encias do Espa\c{c}o, Faculdade de
Ci\^encias da Universidade de Lisboa, Edif\'icio C8, Campo Grande,
P-1749-016, Lisbon, Portugal}
\author{Hooman Moradpour\orcidHooman\!\!}
\email{hn.moradpour@maragheh.ac.ir}
\affiliation{Research Institute for Astronomy and Astrophysics of Maragha (RIAAM),
University of Maragheh, P.O. Box 55136-553, Maragheh, Iran}
\date{\today}

\begin{abstract}
We consider the evolution of traversable wormhole geometries
in the inflationary, radiation-- and matter--dominated eras, and
dynamic wormholes with a traceless energy-momentum tensor (EMT),
within the recently proposed {\it pole dark energy} model. We show
that the evolving radiation-- and matter--dominated wormhole
spacetimes satisfy the null energy condition (NEC), but possess negative energy densities
at late times, thus violating the weak energy condition (WEC)
in this specific domain. 
However, we demonstrate with a specific example that the traceless EMT evolving wormholes, supported by conformally invariant massless fields, in principle, could satisfy the WEC, and consequently the NEC, at all times and for all values of the radial coordinate. Thus, one may imagine a scenario in which these geometries originate in the Planckian era through quantum gravitational processes. Inflation could then provide a natural mechanism for the enlargement of these Planckian wormholes, where their FLRW background evolution is governed by pole dark energy. For the first time in the literature, specific dynamical $4$-dimensional solutions are presented that satisfy the NEC and WEC everywhere and everywhen.

\end{abstract}

\maketitle

\section{Introduction}
General relativistic traversable wormholes as cosmic compact objects, and theoretically
engineered as hypothetical short-cuts in spacetime \cite{MTwormhole,Morris:1988tu}, are threaded and sustained by exotic matter, which is a fluid that violates the null energy condition (NEC).
While an extensive variety of wormhole structures have been explored in the contexts of general relativity and its alternative theories from different aspects \cite{Visser:1995cc,Lobo:2017oab,Antoniou:2019awm, Tangphati:2019pxh, Papantonopoulos:2019ugr, Godani:2019kgy, Banerjee:2020uyi, Fayyaz:2020jzh, Restuccia:2020wls, Korolev:2020ohi,Lazov:2017tjs,Savelova:2019lye, Bak:2019nnu, Xu:2020wfm, Jusufi:2020rpw, Berry:2020tky, Fallows:2020ugr, Maldacena:2020sxe},
evolving wormholes under the effect of cosmic fluids, such as dark
energy and radiation fields, are a relatively outstanding interesting topic.
One way to study this subject is to embed a wormhole in a 
Friedmann-Lema\^{i}tre-Robertson-Walker (FLRW) metric, which permits the geometry to evolve in a cosmological background
\cite{Roman:1992xj,9710026,0608003,0905.3882,0905.4116,bordriazi2011,sajriazi2011,1307.4122,1612.05077,
1406.5703,Cataldo:2008pm,Cataldo:2008ku,Cataldo:2012pw,Harada:2007tj,Maeda:2007tk,Maeda:2009tk}.
A further advantage of these evolving
wormholes, as compared to their static counterparts, is their
ability to satisfy the energy conditions in arbitrary finite
intervals of time \cite{ec0,ec1}.

The line element of an evolving wormhole, used throughout this work, is
given by
\begin{equation}
ds^{2}=-e^{2\Phi (r) }dt^{2}+a^2(t)\left[ \frac{%
dr^{2}}{1-b(r) /r}+r^{2}d\Omega ^{2}\right] ,  \label{met}
\end{equation}%
where $d\Omega ^{2}=d\theta ^{2}+\sin^2 \theta \,d\varphi ^{2}$ is the
linear element of the unit sphere, and $\Phi \left( r\right)$, $b\left(
r\right) $ and $a\left( t\right) $ are the redshift and shape
functions and the scale factor, respectively. In order to describe a
wormhole, the following conditions need to be satisfied: $b(r_{0})=r_{0}$, $1-b(r)/r\geq 0$ and $b^{\prime }(r)-b\left( r\right)/r<0$,
where $r_{0}$ is the wormhole throat, which represents a minimum
radius in the wormhole spacetime \cite{MTwormhole}. The last
inequality translates the flaring-out condition, and through the
Einstein field equations, it imposes the violation
of the NEC
\cite{MTwormhole,Morris:1988tu,Visser:1995cc,Lobo:2017oab}. As the
violation of the energy conditions is a somewhat problematic
issue, it is important to minimize these violations
\cite{1406.5703,1501.04773,1506.03427,1510.07089,Harko:2013yb,Capozziello:2013vna,Capozziello:2014bqa}.

In this paper, we study the evolution of traversable wormholes in
a FLRW background within the recently proposed {\it pole dark
energy} model \cite{1911.01606}. In this model, used to explain dark energy, the
Lagrangian
is the summation of the potential $V$ and a kinetic term of the form $%
-k\left( \nabla \sigma \right) ^{2}/2\sigma ^{p}$ with a pole of
order $p$ and residue $k$ at $\sigma =0$, and thus, the
$p=2$ and $V=0$ case corresponds to a minimal $k$-essence model up to the first order of approximation
\cite{k1}. The kinetic term can be transformed to a canonical scalar field
form, where the resultant transformed
Lagrangian of the model could give rise to an observationally viable
dark energy equation of state evolution, given by $\omega \left(
z\right) <-0.9$, an outcome which occurs even for transformed
potentials $V\left( \phi \right) $ with the forms that could not
normally produce a reliable behavior for the dark energy equation
of state \cite{1911.01606}. Due to their quantum stability and
attractor features, these models with kinetic terms including a
pole have been employed for studying inflation.
A multipole dark energy model has also been proposed \cite{1912.10830}.
Here, we explore the possibility that evolving wormhole geometries may be
supported by this model, in a manner analogous to more standard equations of
state \cite%
{Sushkov:2005kj,Lobo:2005us,Lobo:2005yv,Lobo:2005vc,Lobo:2006ue,DeBenedictis:2008qm,Lobo:2012qq}.
Furthermore, we explore the energy conditions for matter threading these traversable
wormhole geometries.

The paper is outlined in the following manner: In Sec. \ref{secII}, we present the action and the field equations of the pole dark energy model. In Sec. \ref{secIII}, we analyse evolving wormholes in the inflationary, radiation-- and matter--dominated eras, as well as wormholes with a traceless EMT, and explore the validity of the null and weak energy conditions for the solutions obtained. Finally, in Sec. \ref{conclusion}, we summarize and discuss our results.

\section{Action and Field Equations}\label{secII}
%
The action of the pole dark energy model is written as
\begin{equation}
S=\int {\mathrm{d}}^{4}x\sqrt{-g}\left( \frac{R}{2\kappa }+\mathcal{L}%
_{\sigma }+\mathcal{L}_{m}\right) ,  \label{act}
\end{equation}%
where $R$ is the scalar curvature, $\kappa $ is related to
Newton's constant, $\mathcal{L}_{m}$ is the matter Lagrangian density and
the Lagrangian of the scalar field $\sigma $ is given by \cite{1911.01606}
\begin{equation}
\mathcal{L}_{\sigma }=-\frac{1}{2}\frac{k}{\sigma ^{p}}\nabla _{\mu }\sigma
\nabla ^{\mu }\sigma -V\left( \sigma \right) ,  \label{Lsig}
\end{equation}%
in which the pole resides at $\sigma =0$ and has a residue $k$
and order $p$.
Varying the action (\ref{act}) with respect to the metric $g_{\mu \nu }$ and
scalar field $\sigma $, yields the field equations
\begin{eqnarray}
G_{\mu \nu } &=&\kappa \left( T_{\mu \nu }^{\sigma }+T_{\mu \nu }^{m}\right)
, \\
0 &=&\frac{1}{2}\frac{kp}{\sigma ^{p+1}}\nabla _{\varsigma }\sigma \nabla
^{\varsigma }\sigma -\frac{k}{\sigma ^{p}}\nabla _{\varsigma }\nabla
^{\varsigma }\sigma +\frac{dV(\sigma )}{d\sigma },
\end{eqnarray}%
respectively, where%
\begin{equation}
T_{\mu \nu }^{\sigma }=\frac{k}{\sigma ^{p}}\nabla _{\mu }\sigma \nabla
_{\nu }\sigma -\frac{1}{2}\frac{k}{\sigma ^{p}}g_{\mu \nu }\nabla
_{\varsigma }\sigma \nabla ^{\varsigma }\sigma -V(\sigma )g_{\mu \nu },
\notag
\end{equation}
is the scalar field energy-momentum tensor (EMT), $G_{\mu \nu }$ is the Einstein
tensor and $T_{\mu \nu }^{m}$ is the matter EMT.
For the wormhole solutions consider here, the matter EMT is given by $T^{\mu \nu}=\rho u^{\mu} u^{\nu}-\tau n^{\mu}_r n^{\nu}_r +p (n^{\mu}_\theta n^{\nu}_\theta+n^{\mu}_\phi n^{\nu}_\phi)$, where $\rho $ is the energy density, $\tau $ is the radial tension (which is equivalent to a negative radial pressure, i.e., $\tau=-p_r$), $p$ is the tangential pressure and $u^\mu$ and $n^{\mu}_i$ are the unit timelike and spacelike vectors, respectively \cite{Lobo:2017oab}. According to this relation, one verifies that $T_{\nu }^{\mu }={\mathrm{diag(}} -\rho ,-\tau , p, p{\mathrm{)}}$.

Note that the kinetic term in Eq. (\ref{Lsig}) can be transformed to a
canonical form $-\left( \nabla \phi \right) ^{2}/2$ via \cite{1911.01606}
\begin{equation}
\sigma =\left\{
\begin{tabular}{lc}
$\left( \frac{|2-p|}{2\sqrt{k}}\right) ^{2/(2-p)}\,\phi ^{2/(2-p)}$ & for $%
p\neq 2$ \\
$e^{\pm \phi /\sqrt{k}}$ & for $p=2$%
\end{tabular}%
\right. \ .  \label{sigphi}
\end{equation}%
With this transformed canonical Lagrangian of the scalar field in hand, the field equations are given by
\begin{eqnarray}
G_{\mu \nu } &=&\kappa \left( T_{\mu \nu }^{\phi }+T_{\mu \nu }^{m}\right) ,
\label{ein} \\
0 &=&-\nabla _{\varsigma }\nabla ^{\varsigma }\phi +\frac{dV(\phi )}{d\phi },
\label{scalar}
\end{eqnarray}%
where $T_{\mu \nu }^{\phi }=\nabla _{\mu }\phi \nabla _{\nu }\phi -\frac{1}{2}%
g_{\mu \nu }\nabla _{\varsigma }\phi \nabla ^{\varsigma }\phi -V(\phi
)g_{\mu \nu }$,
with $V(\phi )=V(\sigma \left( \phi \right) )$, and $\sigma \left( \phi
\right) $ can be read from Eq. (\ref{sigphi}).

Eq. (\ref{ein}) may be expressed as the following effective Einstein field equation, $G_{\mu \nu }=\kappa T^{\rm eff}_{\mu \nu }$, where the effective EMT is given by $T^{\rm eff}_{\mu \nu }=  T_{\mu \nu }^{\phi }+T_{\mu \nu }^{m}$. The necessary condition to have a wormhole geometry is the violation of the generalized NEC \cite{Harko:2013yb}, i.e., $T^{\rm eff}_{\mu \nu } k^\mu k^\nu<0$ \cite{Harko:2013yb,Capozziello:2013vna,Capozziello:2014bqa}. Indeed, one may, in principle, impose that the matter EMT satisfies the NEC, i.e., $T^m_{\mu \nu } k^\mu k^\nu \geq 0$ and thus, the pole dark energy plays the role of the exotic matter in order to support the geometry. More specifically, taken into account the above considerations, the condition $T^{\rm eff}_{\mu \nu } k^\mu k^\nu<0$ implies $0 \leq T^m_{\mu \nu } k^\mu k^\nu < - T_{\mu \nu }^{\phi } k^\mu k^\nu$. We show below that this is indeed possible, and consequently we construct specific dynamical $4$-dimensional solutions that satisfy the NEC everywhere and everywhen.

It is also interesting to note the role played by the scale factor, given in the line element of an evolving geometry (\ref{met}), in describing wormholes and the satisfaction of the wormhole conditions \cite{Roman:1992xj}. To this effect, in order to verify that the ``wormhole'' form of the metric is preserved with time, we consider an embedding of a $t={\rm const}$ and an equatorial slice $\theta=\pi/2$ of the spacetime given by Eq. (\ref{met}), in a flat 3-dimensional Euclidean space with metric
\begin{equation}
ds^2=d{\bar{z}}^2+d{\bar{r}}^2+{\bar{r}}^2\,{d\varphi}^2\,.
\label{barredslice}
\end{equation}
Here, the wormhole slice is given by the following metric
\begin{equation}\label{slice}
ds^2={a^2(t)\,{dr^2}\over{1-b(r)/r}} + a^2(t)\,
r^2\,d\varphi^2\,,
\end{equation}
and confronting the coefficients of ${d\varphi}^2$, provides the following relations
\begin{eqnarray}
\bar{r}&=&{a(t)\,r}\big|_{t={\rm const}} \,,
       \label{coef1:phi}      \\
{d\bar{r}}^2&=&a^2(t)\,{dr}^2\big|_{t={\rm const}}  \,.
\label{coef2:phi}
\end{eqnarray}
We emphasize, in particular, that when considering derivatives, that Eqs. (\ref{coef1:phi}) and (\ref{coef2:phi}) do not represent a ``coordinate transformation'', but rather a ``rescaling'' of the
$r$ coordinate on each $t={\rm constant}$ slice \cite{Roman:1992xj}.

With respect to the ${\bar{z}},{\bar{r}},\varphi$ coordinates, the ``wormhole'' form of the metric will be preserved if the metric on the embedded slice has the form
\begin{equation}\label{WHslice}
ds^2={{d{\bar{r}}^2}\over{1-{\bar{b}(\bar{r})/{\bar{r}}}}} +
                      {\bar{r}}^2{d\varphi}^2\,,
\end{equation}
where $\bar{b}(\bar{r})$ has a minimum at some $\bar{b}(\bar{r}_0)=\bar{r}_0$. Eq. (\ref{slice}) can be rewritten in the form of Eq. (\ref{WHslice}) by using Eqs. (\ref{coef1:phi}) and (\ref{coef2:phi}) and $\bar{b}(\bar{r})=a(t)\,b(r)$.
The evolving wormhole will have the same overall size and shape relative to the ${\bar{z}},{\bar{r}},\varphi$ coordinate system, as the initial wormhole had relative to the initial $z,r,\varphi$ embedding space coordinate system. This is due to the fact that the embedding space corresponds to $z,r$ coordinates which ``scale'' with time (each embedding space corresponds to a particular value of $t={\rm constant}$). 

Following the embedding procedure outlined in Ref. \cite{MTwormhole}, and using Eqs. (\ref{barredslice}) and (\ref{WHslice}), one deduces that
\begin{equation}
{{d{\bar{z}}}\over{d{\bar{r}}}}
=\pm\left({{\bar{r}}\over{\bar{b}(\bar{r})}}-1\right)^{-1/2}
={{dz}\over{dr}} \,.
            \label{barredembedding}
\end{equation}
Eq. (\ref{barredembedding}) implies
\begin{eqnarray}
\bar{z}(\bar{r}) &=& \pm\int{{d\bar{r}}\over{(\bar{r}/{\bar{b}(\bar
{r})}-1)^{1/2}}}
	\nonumber \\
&=& \pm \, a(t)\,\int{\left(r/b-1\right)^{-1/2}} \,dr
	\nonumber \\
&=& \pm \,a(t)\,z(r)\,.
            \label{embed:relation}
\end{eqnarray}
Thus, taking into account Eqs. (\ref{coef2:phi}) and (\ref{embed:relation}), we verify that the relation between the embedding space at any time $t$ and the initial embedding space at $t=0$ is given by
\begin{equation}
ds^2=d{\bar{z}}^2+d{\bar{r}}^2+{\bar{r}}^2\,{d\varphi}^2
        =a^2(t)\,\left(dz^2+dr^2+r^2{d\varphi}^2\right).
\end{equation}
Relative to the ${\bar{z}},{\bar{r}},\varphi$ coordinate system the wormhole maintains the same size, as the scaling of the embedding space compensates for the evolution of the wormhole.
However, the wormhole will change size relative to the initial $t=0$ embedding space.

Writing the analog of the ``flaring out condition'' \cite{MTwormhole} for the evolving wormhole we have $d\,^2{\bar{r}(\bar{z})}/d{\bar{z}}^2>0$ at or near the throat. Thus, taking into account the above expressions, we have
\begin{equation}
{{d\,^2{\bar{r}(\bar{z})}}\over{d{\bar{z}}^2}}
       =\frac{1}{a(t)}\,{{b-b'r}\over{2b^2}}
       =\frac{1}{a(t)}\,{{d\,^2r(z)}\over{dz^2}}>0\,,
       \label{barred:flareout}
\end{equation}
at or near the throat. We also deduce the expressions ${\bar{b}}'(\bar{r})={{d\bar{b}}/{d\bar{r}}}=b'(r)={{db}/{dr}}$, so that one may rewrite the right-hand side of Eq. (\ref{barred:flareout})
relative to the barred coordinates as
\begin{equation}
{{d\,^2{\bar{r}(\bar{z})}}\over{d{\bar{z}}^2}}
=\left({{\bar{b}-{\bar{b}}'\bar{r}}\over{2{\bar{b}}^2}}\right)>0\,,
         \label{barred:flareout2}
\end{equation}
at or near the throat. One verifies that using the barred coordinates, the flaring out condition Eq.
(\ref{barred:flareout2}), has the same form as for the static wormhole.

In this paper, we consider a specific class of wormhole solutions with a
constant redshift function, $\Phi =\mathrm{const}$.
Using the metric (\ref{met}), the
gravitational field equations (\ref{ein}) provide
\begin{eqnarray}
\rho (t,r) &=&\rho _{b}\left( t\right) -\rho _{\phi }\left( t\right) +\frac{%
b^{\prime }}{r^{2}a^{2}},  \label{rhoeq} \\
\tau \left( t,r\right) &=&\tau _{b}\left( t\right) -\tau _{\phi }\left(
t\right) +\frac{b}{r^{3}a^{2}}, \\
p\left( t,r\right) &=&-\tau _{b}\left( t\right) +\tau _{\phi }\left(
t\right) -\frac{b^{\prime }}{2r^{2}a^{2}}+\frac{b}{2r^{3}a^{2}},
\label{pteq}
\end{eqnarray}%
where $\rho _{b}\left( t\right) =3H^{2}$, $\tau _{b}\left( t\right) =H^{2}+2\ddot{a}/a$,
$\rho _{\phi }\left( t\right) =\dot{\phi}^{2}/2+V(\phi )$, and $\tau _{\phi
}\left( t\right) =-\dot{\phi}^{2}/2+V(\phi )$,
in which $H=\dot{a}/a$. Here, the overdot and prime denote derivatives with
respect to $t $ and $r$, respectively. For notational simplicity, we
consider $\kappa =1$. Note that if one fixes $a$ to unity and excludes the
background evolution and the dark energy contribution, we recover the
well-known equations of motion of the Morris-Thorne wormhole \cite{MTwormhole}.

From Eq. (\ref{scalar}), the scalar field equation of motion is given by
$\ddot{\phi}+3H\dot{\phi}+dV/d\phi=0$,
%
with $\phi =\phi \left( t\right) $. In order to solve this equation
numerically, we re-write it in terms of dimensionless functions of $a$. To
this effect, we use the following definitions: $\ddot{\phi}=\ddot{a}\phi ^{\prime }(a)+\dot{a}^{2}\phi ^{\prime \prime}(a)$, $\dot{\phi}=\dot{a}\phi ^{\prime }(a)$, $\ddot{a}=H\dot{a}+\dot{H}a$, $\dot{H}=\dot{a}H^{\prime }(a)$, and $\dot{a}=Ha$,
where here the prime denotes a derivative with respect to the scale factor.
We also define $U=V/3H_{0}^{2}$ and $E=H/H_{0}$, where $H_{0}$ is the present
value of the Hubble parameter. Thus, we obtain the following differential equation:
\begin{equation}
\phi ^{\prime \prime }(a)a^{2}E^2(a)+\phi ^{\prime }(a)aE(a)\left[
4E(a)+aE^{\prime }(a)\right] +3\frac{dU}{d\phi }=0.  \label{phide}
\end{equation}%
In addition to this, Eqs. (\ref{rhoeq})-(\ref{pteq}) can be re-expressed as
\begin{eqnarray}
\frac{\rho}{3 H_0^2} &=& \frac{b'(r)}{3 a^2 H_0^2 r^2}-\frac{1}{6} a^2 E^2(a) \phi'^2(a)
	\nonumber \\
&&-U \left(\phi \right)+E^2(a) , \label{rhodimless} \\
\frac{\tau}{3 H_0^2} &=& \frac{b(r)}{3 a^2 H_0^2 r^3}+\frac{1}{6} a^2 E^2(a) \phi'^2(a)
	\nonumber \\
&&-U \left(\phi \right)+\frac{2}{3} a E(a) E'(a)+ E^2(a), \\
\frac{p}{3 H_0^2} &=& -\frac{b'(r)}{6 a^2 H_0^2 r^2}+\frac{b(r)}{6 a^2 H_0^2 r^3}-\frac{1}{6} a^2 E^2(a) \phi'^2(a)
		\nonumber \\
&&+U \left(\phi \right)-\frac{2}{3} a E(a) E'(a)-E^2(a), \label{ptdimless}
\end{eqnarray}%

To solve Eq. (\ref{phide}) for $\phi \left( a\right) $, we
have to deduce $E$. By applying a barotropic equation of state $\tau _{b}=-\omega
_{b}\rho _{b}$ for the background, we find $E=a^{-3\left( \omega
_{b}+1\right) /2}$. For the inflationary, radiation-- and matter--dominated
eras, the parameter $\omega _{b}$ is equal to $-1$, $1/3$ and $0$,
respectively.

\begin{figure*}[htb!]
\centering%
\subfigure[~$1-b(r)/r$] {\includegraphics[width=.35\textwidth]{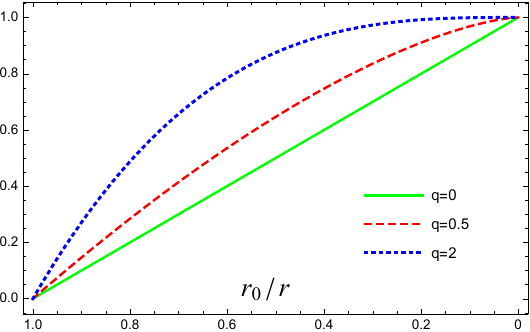} \qquad}
\subfigure[~$b'(r)-b(r)/r$] {\includegraphics[width=.35\textwidth]{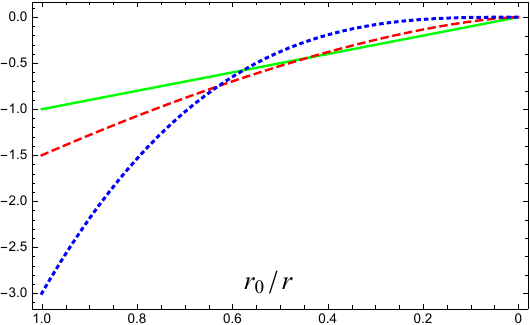}
}
\caption{The conditions $1-b(r)/r \geq 0$ and $b^{\prime }(r)-b\left( r\right)/r<0$ for the shape function
$b(r)=r_{0}\left( r_{0}/r\right) ^{q}$ with $q=0$, $0.5$ and $2$.}
\label{shapefunc}
\end{figure*}

Relative to the potential $V\left( \phi \right) $, from Eq. (\ref%
{sigphi}), we see that a power law potential $V\sim \sigma ^{n}$
transforms to another power law potential of the form $\phi ^{2n/(2-p)}$.
For $p<2$, the signs of the initial and transformed potential powers are the
same while the transformed one is steeper and so is less interesting for
inflation or dark energy close to a cosmological constant like behavior. For
$p>2$, the signs flip, i.e., a monomial potential is transformed to an
inverse power law one and vice versa. This is significant as for
canonical scalar fields, a monomial potential causes a thawing dark energy
scenario which begins with a cosmological constant like state at high
redshift and deviates from this as it evolves at later times. On the other
hand, an inverse power law potential exhibits freezing dark energy behavior
at early times, i.e., it could possess a dynamical attractor behavior with a
constant equation of state parameter $\omega _{\phi }=-\tau _{\phi }/\rho
_{\phi }$ and then advances towards a cosmological constant behavior at
later times \cite{caldlin}. Therefore, the pole dark energy model can
produce the features of freezing, possibly attractor, fields from monomial
potentials and thawing fields from an initial inverse power law potential.
Here, we use the power law potential for $\sigma $ with $n>0$ and $p>2$
which causes an inverse power law potential for $\phi $ of the form $V\sim
\phi ^{-\alpha }$ in which $\alpha =2n/(p-2)$. Note that any value of $%
\alpha $ could be obtained by different sets of $n(>0)$ and  $p(>2)$.

In order to study the behavior of the EMT components $\rho $, $%
\tau $ and $p$ given by Eqs. (\ref{rhoeq})-(\ref{pteq}) and the
corresponding energy conditions, we have to choose a suitable shape function
$b(r)$ for our wormhole structure. For this purpose, we choose $%
b(r)=r_{0}\left( r_{0}/r\right) ^{q}$ where $r_{0}$ is the wormhole throat and $b(r_{0})=r_{0}$.
As mentioned in the Introduction, the shape function satisfies $1-b(r)/r \geq 0$ and the flaring-out condition $b^{\prime }(r)-b\left( r\right)/r<0$. These conditions lead to $1-b(r)/r=1-\left(r_0/r\right)^{1+q} \geq 0$ and $b^{\prime}(r)-b\left( r\right)/r=-(1+q)\left(r_0/r\right)^{1+q}<0$, respectively. As one can see, both of them are satisfied provided $q>-1$ (Note that $r_0/r \leq 1$). For specific values of $q$ used in the following analysis, we show the satisfaction of these conditions for the shape function in Fig. \ref{shapefunc}.

\section{Evolving traversable wormholes and energy conditions}\label{secIII}

\begin{figure*}[htb!]
\centering%
\subfigure[~Radiation-dominated era ($\omega_b=1/3$) with
$U(\phi)=\phi^{0.1}$] {\label{fig1a}\includegraphics[width=.322\textwidth]{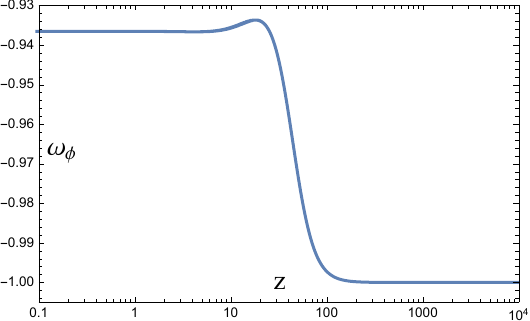} }
\subfigure[~Matter-dominated era ($\omega_b=0$) with $U(\phi)=\phi^{0.2}$] {\label{fig1b}\includegraphics[width=.322\textwidth]{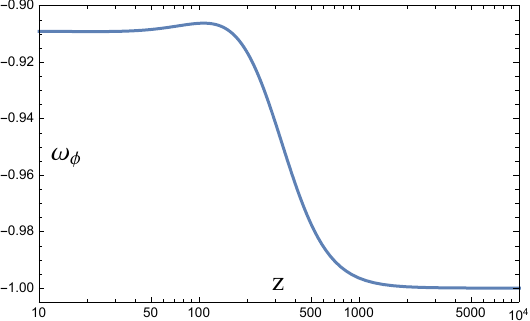}
} \subfigure[~Traceless EMT case with $U(\phi)=\phi^{0.1}$]
{\label{fig1c}\includegraphics[width=.322\textwidth]{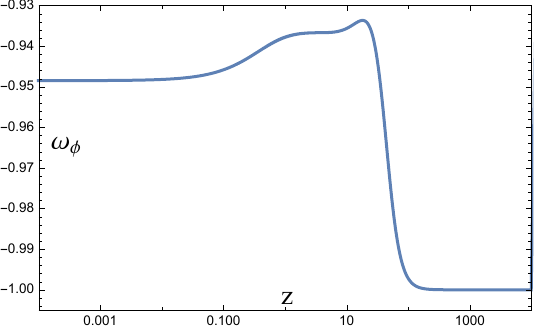} }
\caption{The behavior of $\protect\omega _{\protect\phi }$ vs $z$ for the
radiation-dominated era, the matter-dominated era and the traceless EMT case.
Note that the horizontal axis is logarithmic. See the text for more details.}
\label{fig1}
\end{figure*}
\begin{figure*}[htb!]
\centering\subfigure[~$\abs{\rho}/3H_0^2$ vs $z$]
{\label{fig2a}\includegraphics[width=.3225\textwidth]{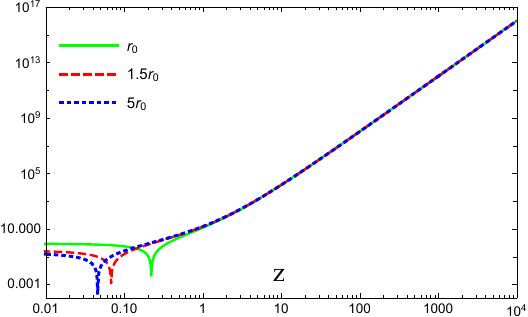}}\subfigure[~$(\rho-\tau)/3H_0^2$ vs $z$] 
{\label{fig2b}\includegraphics[width=.3225\textwidth]{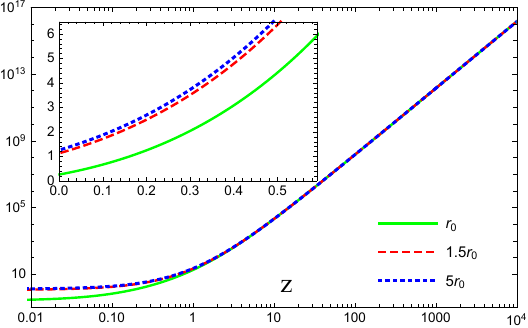}
} \subfigure[~$(\rho+p)/3H_0^2$ vs $z$]
{\label{fig2c}\includegraphics[width=.3225\textwidth]{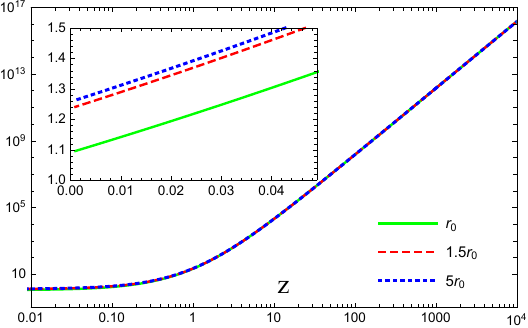} }
\caption{The behaviors of $\protect\rho $, $\protect\rho -\protect\tau $ and
$\protect\rho +p$, respectively, versus $z$ for different values of $r$ in
the radiation-dominated era ($\protect\omega_b=1/3$) with $U(\protect\phi)=%
\protect\phi^{0.1}$ and $q=2$. Note that both the horizontal and vertical axes are logarithmic. The $\gamma$-shaped part in subfigure (a) shows the point at which $\rho$ meets zero and changes its sign.}
\label{fig2}
\end{figure*}
%
%
%
\begin{figure*}[htb!]
\centering\subfigure[~$\abs{\rho}/3H_0^2$ vs $z$]
{\label{fig3a}\includegraphics[width=.3225\textwidth]{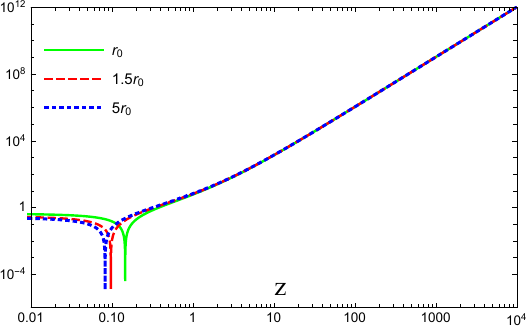} }\subfigure[~$(\rho-\tau)/3H_0^2$ vs $z$] 
{\label{fig3b}\includegraphics[width=.3225\textwidth]{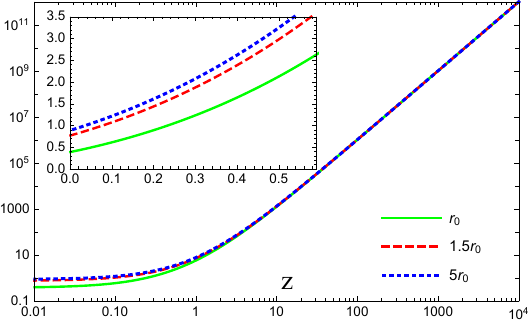}
}\subfigure[~$(\rho+p)/3H_0^2$ vs $z$]
{\label{fig3c}\includegraphics[width=.3225\textwidth]{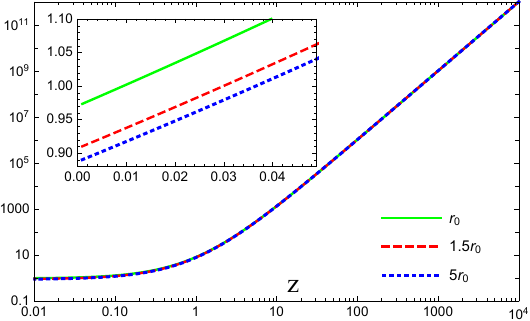} }
\caption{The behaviors of $\protect\rho $, $\protect\rho -\protect\tau $ and
$\protect\rho +p$, respectively, versus $z$ for different values of $r$ in
the matter-dominated era ($\protect\omega_b=0$) with $U(\protect\phi)=%
\protect\phi^{0.2}$ and $q=0.5$. Both horizontal and vertical axes are logarithmic. The $\gamma$-shaped part in subfigure (a) shows the point at which $\rho$ attains zero and consequently changes sign.}
\label{fig3}
\end{figure*}
\begin{figure*}[htb!]
\centering\subfigure[~$\rho/3H_0^2$ vs $z$]
{\label{fig4a}
\includegraphics[width=.323\textwidth]{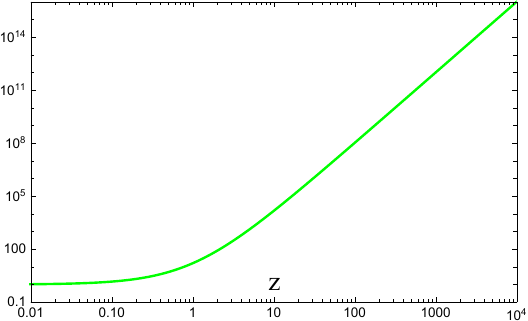}}\subfigure[~$(\rho-\tau)/3H_0^2$ vs $z$]
{\label{fig4b}\includegraphics[width=.323\textwidth]{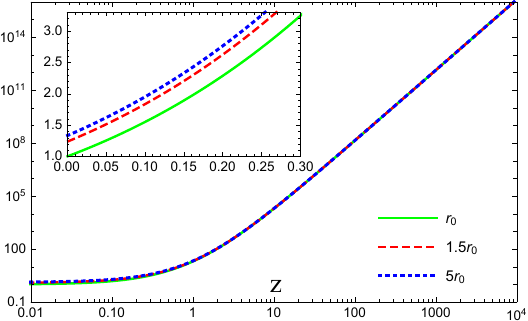}
}\subfigure[~$(\rho+p)/3H_0^2$ vs $z$]{\label{fig4c}
\includegraphics[width=.323\textwidth]{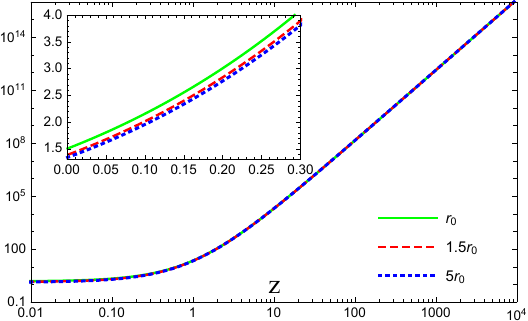} }
\caption{The behaviors of $\protect\rho $, $\protect\rho -\protect\tau $ and
$\protect\rho +p$, respectively, versus $z$ for different values of $r$ for
a traceless EMT with $U(\protect\phi)=\protect\phi^{0.1}$
and $q=0$. Note that both horizontal and vertical axes are logarithmic.}
\label{fig4}
\end{figure*}
In this section, we study the evolution of traversable wormholes
in the inflationary, radiation-- and matter--dominated eras, as
well as evolving wormholes with a traceless EMT, in the presence
of pole dark energy. We will also explore the null and weak energy
conditions for our solutions. The weak energy condition (WEC) is
expressed in terms of the energy density $\rho $, radial tension
$\tau $ and tangential pressure $p$ as
$\rho \geq 0$, $\rho -\tau \geq 0$ and $\rho +p\geq 0$,
respectively. The last two inequalities, i.e., $\rho -\tau \geq 0$
and $\rho +p\geq 0$ correspond to the NEC.

In the following, we consider $U(\phi )=\phi ^{-\alpha }$ and $b(r)=r_{0}\left(
r_{0}/r\right) ^{q}$. Then, with $E=a^{-3\left( \omega _{b}+1\right) /2}$
which arises from the background equation of state $\tau _{b}=-\omega _{b}\rho
_{b}$ and using the dimensionless definitions,  Eqs. (\ref{rhoeq})-(\ref{pteq})
lead to
\begin{equation}
\frac{\rho }{3H_{0}^{2}} = \frac{6-a^{2}\phi'^2(a)}{a^{3(\omega _{b}+1)}}
-\phi^{-\alpha}(a)-\frac{q r_{0}\left( \frac{r_{0}}{r}\right) ^{q}}{3a^{2}H_{0}^{2}r^{3}},
\label{rho}
\end{equation}%
\begin{equation}
\frac{\rho -\tau }{3H_{0}^{2}} =\frac{3(1+\omega _{b})-a^{2}\phi'^2(a)}{a^{3(\omega _{b}+1)}}
-\frac{\left( q+1\right) r_{0}\left( \frac{r_{0}}{r}\right) ^{q} }{%
3a^{2}H_{0}^{2}r^{3}},
  \label{rhotau}
\end{equation}
\begin{equation}
\frac{\rho +p}{3H_{0}^{2}} =\frac{3(1+\omega _{b})-a^{2}\phi'^2(a)}{a^{3(\omega _{b}+1)}}
-\frac{\left( q-1\right) r_{0}\left( \frac{r_{0}}{r}\right) ^{q} }{%
6a^{2}H_{0}^{2}r^{3}},
\end{equation}%
respectively.
In order to keep the equations independent of $H_0$, we will consider the wormhole throat as $r_0 = A H_0^{-1}$, where $A$ is a dimensionless constant, and compute the equations numerically for $r=B r_0=A B H_0^{-1}$ where $B$ is also an arbitrary dimensionless constant which could vary from $1$ to infinity in order to cover all radii $r \geq r_0$. In what follows, we set $A$ to unity.
We also consider that the initial conditions for solving Eq. (\ref{phide})
numerically are $\phi (\epsilon )=\phi ^{\prime }(\epsilon )=10^{-4}$, where $%
\epsilon $ is very close to $a=0$.

\subsection{Inflation era}

%
For the inflationary era with $\omega _{b}=-1$, Eq. (\ref{rhotau}) reduces to%
\begin{equation*}
\frac{\rho -\tau }{3H_{0}^{2}}=-\frac{a^{2}}{3}\phi'^{2}(a)-\frac{r_{0}\left( \frac{r_{0}}{r}\right) ^{q}\left( q+1\right)
}{3a^{2}H_{0}^{2}r^{3}},
\end{equation*}%
which immediately leads to $\rho -\tau <0$ for $q>-1$, which satisfies the flaring-out  condition at the throat. Thus, both the NEC and the WEC are violated, if
one seeks for a traversable wormhole in this region. It is, however, remarkable
that $\omega _{\phi }$ is physically viable for this case, i.e., $\omega
_{\phi }\left( z\right) <-0.9$, with some $\alpha$ values less than unity, according to our numerical analysis.

\subsection{Radiation-dominated era}

%
The behavior of the equation of state $\omega _{\phi }=-\tau _{\phi }/\rho
_{\phi }=(\dot{\phi}^{2}/2-V(\phi ))/(\dot{\phi}^{2}/2+V(\phi ))$ with
respect to the redshift $z$ $(=1/a-1)$ in the radiation-dominated era where $%
\omega _{b}=-\tau _{b}/\rho _{b}=1/3$, $E=a^{-2}$ and $a(t) \propto t^{1/2}$ with $U\left( \phi
\right) =V\left( \phi \right) /3H_{0}^{2}=\phi ^{0.1}$ is depicted in Fig. %
\ref{fig1a}.
It is physically viable since $\omega _{\phi }\left( z\right)
<-0.9$. In Fig. \ref{fig2}, the behaviors of $\rho $, $\rho -\tau $ and $%
\rho +p$ versus $z$ for different values of $r$ are shown where $q=2$, i.e.,
$b(r)=r_{0}\left( r_{0}/r\right) ^{2}$, which satisfies all required conditions.
As one can see, at earlier times, the wormhole geometry satisfies the WEC. As time
passess, $\rho $, $\rho -\tau $ and $\rho +p$ decrease. This occurs for the
throat as well as other wormhole radii. Eventually, at late times, the energy
density $\rho $ becomes negative, as depicted in Fig. \ref{fig2a}, whereas $\rho
-\tau $ and $\rho +p$ remain positive (see Figs. \ref{fig2b} and \ref{fig2c}).
Thus, the NEC is satisfied at late times, contrary to the WEC. Consequently,
the NEC is satisfied by these evolving traversable wormhole solutions at all
times and for all values of $r$, including the wormhole throat. Note that the energy
density of the throat becomes negative earlier than other radii, as is transparent from Fig. \ref{fig2a}.

\subsection{Matter-dominated era}

%
The behavior of $\omega _{\phi }$ versus $z$ in the matter-dominated era
where $\omega _{b}=0$, $E=a^{-3/2}$ and $a(t) \propto t^{2/3}$ with $U\left( \phi \right) =\phi
^{0.2}$ is depicted in Fig. \ref{fig1b}. It is also physically viable since $%
\omega _{\phi }\left( z\right) <-0.9$. In Fig. \ref{fig3}, the behaviors of $%
\rho $, $\rho -\tau $ and $\rho +p$ with respect to $z$ for
different values of $r$ are shown for $q=0.5$, i.e., $b(r)=r_{0}\left(
r_{0}/r\right) ^{0.5}$ which satisfies all the required conditions.
As exhibited in Fig. \ref{fig3}, the wormhole geometry satisfies the WEC at earlier times, and as it evolves in time,
the quantities $\rho $, $\rho -\tau $ and $\rho +p$ decrease. The throat and the other wormhole
radii behave in this manner, and at late times, the energy
density $\rho $ becomes negative. However, the quantities $\rho -\tau $ and $\rho +p$ remain
positive at late times (Figs. \ref{fig3b} and \ref{fig3c}).
Therefore, the NEC is satisfied by these dynamical wormhole solutions at all times
for all values of $r$, including the wormhole throat, as in the previous example.
However, the WEC is violated only at late times, as depicted by Fig. \ref{fig3a}. In addition to this,
the energy density of the throat becomes negative earlier than for regions of larger radii.

\subsection{Wormholes with traceless EMT}

%
Considering the traceless EMT, i.e., $-\rho -\tau +2p=0$, we obtain from Eqs. (\ref{rhodimless})-(\ref{ptdimless}) that
\begin{eqnarray}
2aE(a)E^{\prime }(a)+\frac{1}{3}E(a)^{2}\left[ 12+a^{2}\phi ^{\prime }\left(
a\right) ^{2}\right] 
	\nonumber  \\
-4U\left( \phi \right)+\frac{2}{3 a^2 H_0^2}\frac{b'(r)}{r^2} =0.
\label{tracelessEMTb}
\end{eqnarray}
The traceless EMT implies a conformally invariant massless field, commonly encountered in the Casimir effect. In fact, conformal symmetry imposes significant constraints on the structure of conformal field theories, where one can relate and unify physical theories.
In order to solve the coupled differential equation system, Eqs. (\ref{phide}) and (\ref{tracelessEMTb}), for $\phi $ and $E$, numerically, Eq. (\ref{tracelessEMTb}) should be independent of $r$. This leads to the imposition $b'(r) = C r^2$ where $C$ is an arbitrary constant, and consequently provides the shape function $b(r)=r_0+C(r^3-r_0^3)/3$, with $b(r_0)=r_0$. This shape function satisfies the required conditions for $C \leq 0$, however, we set $C$ to zero, which is equivalent to the $q=0$ case for the shape function $b(r)=r_0$. We also set the initial condition $E\left( \epsilon \right) =10^{12}$. The behavior of $\omega _{\phi }$ versus $z$ with $U\left( \phi \right) =\phi
^{0.1}$ is depicted in Fig. \ref{fig1c}.
It is physically viable since $\omega _{\phi }\left( z\right)
<-0.9$. In Fig. \ref{fig4}, the behaviors of $\rho $, $\rho -\tau $ and $%
\rho +p$, with respect to $z$ for different values of $r$ are shown.
Note that $\rho$ is independent of the radial coordinate $r$ for the $q=0$ case, as can be seen from Eq. (\ref{rho}).
Figure \ref{fig4} shows that $\rho $, $\rho -\tau $ and $\rho +p$ decrease as time evolves. However, they
remain positive at all times and consequently the NEC and WEC are always
satisfied. This occurs for the wormhole throat as well as other wormhole radii.
It is interesting to note that at a specified time/redshift, the quantity $\rho -\tau $ increases for increasing values of the radius, and the minimum value corresponds to the throat, as depicted by Fig. \ref{fig4b}. On the other hand, $\rho+p$ decreases for increasing values of the radius, and has a maximum at the throat, as is depicted by Fig. \ref{fig4c}.

\section{Discussion and Conclusion}\label{conclusion}

%
In this work, using the recently proposed pole dark
energy model, we explored the evolution of traversable wormhole
geometries in a FLRW background, in particular, in the
inflationary, radiation-- and matter--dominated eras. In addition
to these solutions, we also analysed dynamic wormholes with a
traceless EMT. A central theme in this
work was the study of the energy conditions, and it was shown
explicitly that the evolving radiation-- and matter--dominated
wormhole spacetimes satisfy the NEC, but possess negative energy
densities at late times, thus violating the WEC in this specific
domain. Nevertheless, inflating traversable wormhole geometries
always violate both the NEC and WEC. On the other hand, it was
shown for a specific example that the traceless EMT evolving wormholes satisfies both the
NEC and WEC at all times.

These solutions can be thought to be embedded in a scenario where
inflation provides a natural mechanism for the enlargement of
submicroscopic Planckian wormholes, that originated via quantum gravitational processes, to macroscopic size. Their subsequent evolution is governed by pole dark energy.
In fact, it was shown that Lorentzian wormholes in a flat de Sitter
background could serve this purpose \cite{Roman:1992xj}.
Subsequent work on evolving wormholes, conformally related to
static Morris-Thorne wormhole geometries were also found to exist
for finite intervals of time, with the EMT satisfying the WEC in specific ranges \cite{ec0,ec1}. The role
of extra compact decaying dimensions have also been dealt with in
the context of simple models involving an exponential inflation
and a Kaluza–Klein type inflationary scenario \cite{ec1}.

Finally, to the best of our knowledge, the evolving traversable
wormhole geometries considered in this work, are the first found
in the literature, in four-dimensions, to present solutions in a cosmological background constructed by normal matter. More specifically, the NEC and WEC are satisfied everywhere and everywhen. Thus, these novel results motivate further work in this interesting branch of research. Work along
these lines is presently underway.


\begin{acknowledgments}
We thank the referee for the constructive comments that helped us to significantly improve the paper.
MKZ thanks Shahid Chamran University of Ahvaz, Iran, for
supporting this work. FSNL acknowledges support from the Funda\c{c}\~{a}o
para a Ci\^{e}ncia e a Tecnologia (FCT) Scientific Employment Stimulus
contract No. CEECIND/04057/2017, and funding from grants No. CERN/FIS-PAR/0037/2019 and No. PTDC/FIS-OUT/29048/2017.
\end{acknowledgments}

\end{document}